\documentclass[sigconf,natbib=true]{acmart}
\usepackage{svg}

\usepackage{xcolor}
\usepackage{amsmath}
\usepackage{amsfonts}
\usepackage{enumitem}
\usepackage{booktabs}
\usepackage{adjustbox}
\usepackage[noend]{algorithmic}
\usepackage[ruled]{algorithm2e}

\DeclareMathOperator*{\argmax}{arg\,max}

\AtBeginDocument{%
	\providecommand\BibTeX{{%
\normalfont B\kern-0.5em{\scshape i\kern-0.25em b}\kern-0.8em\TeX}}}

\copyrightyear{2024}
\acmYear{2024}
\setcopyright{rightsretained}
\acmConference[SIGIR '24]{Proceedings of the 47th International ACM SIGIR Conference on Research and Development in Information Retrieval}{July 14--18, 2024}{Washington, DC, USA}
\acmBooktitle{Proceedings of the 47th International ACM SIGIR Conference on Research and Development in Information Retrieval (SIGIR '24), July 14--18, 2024, Washington, DC, USA}
\acmDOI{10.1145/3626772.3657769}
\acmISBN{979-8-4007-0431-4/24/07}

\newcommand{\msmarco}{\textsc{Ms Marco}\xspace}
\newcommand{\nq}{\textsc{NQ}\xspace}
\newcommand{\beir}{\textsc{Beir}\xspace}
\newcommand{\quora}{\textsc{Quora}\xspace}

\newcommand{\our}{\textsc{Seismic}\xspace}
\newcommand{\pisa}{\textsc{Pisa}\xspace}
\newcommand{\ioqp}{\textsc{Ioqp}\xspace}
\newcommand{\grassrma}{\textsc{GrassRMA}\xspace}
\newcommand{\pyann}{\textsc{PyAnn}\xspace}
\newcommand{\bruchetal}{\textsc{SparseIvf}\xspace}

\newcommand{\sinnamon}{\textsc{Sinnamon}$_\textsc{Weak}$\xspace}

\newcommand{\splade}{\textsc{Splade}\xspace}
\newcommand{\esplade}{\textsc{E-Splade}\xspace}

\newcommand{\unicoil}{\textsc{uniCoil-T5}\xspace}

\newcommand{\cut}{\textsf{cut}}
\newcommand{\heapfactor}{\textsf{heap\_factor}}
\newcommand{\heap}{\textsc{Heap}}

\newcommand{\etal}{\emph{et al.}\xspace}

\begin{document}

\title{Efficient Inverted Indexes for Approximate Retrieval over Learned Sparse Representations}

\author{Sebastian Bruch}
\affiliation{%
    \institution{Pinecone}
    \city{New York}
    \country{USA}
}
\email{sbruch@acm.org}

\author{Franco Maria Nardini}
\affiliation{%
    \institution{ISTI-CNR}
    \city{Pisa}
    \country{Italy}
}
\email{francomaria.nardini@isti.cnr.it}

\author{Cosimo Rulli}
\affiliation{%
    \institution{ISTI-CNR}
    \city{Pisa}
    \country{Italy}
}
\email{cosimo.rulli@isti.cnr.it}

\author{Rossano Venturini}
\affiliation{%
    \institution{University of Pisa}
    \city{Pisa}
    \country{Italy}
}
\email{rossano.venturini@unipi.it}


\begin{abstract}
Learned sparse representations form an attractive class of contextual
embeddings for text retrieval. That is so because they are effective models
of relevance and are interpretable by design. Despite their apparent compatibility
with inverted indexes, however, retrieval over sparse embeddings remains challenging.
That is due to the distributional differences between learned embeddings and
term frequency-based lexical models of relevance such as BM25.
Recognizing this challenge, a great deal of research has gone into, among other things,
designing retrieval algorithms tailored to the properties of learned sparse representations,
including \emph{approximate} retrieval systems. In fact, this task featured prominently
in the latest BigANN Challenge at NeurIPS 2023, where approximate algorithms were evaluated on a large benchmark
dataset by throughput and recall. In this work, we propose a novel organization of the
inverted index that enables fast yet effective approximate retrieval over learned sparse embeddings.
Our approach organizes inverted lists into geometrically-cohesive blocks, each equipped with
a summary vector. During query processing, we quickly determine if a block must be evaluated
using the summaries. As we show experimentally, single-threaded query processing using our method, \our,
reaches sub-millisecond per-query latency on various sparse embeddings of the \msmarco dataset
while maintaining high recall. Our results indicate that \our is one to two orders of magnitude faster than state-of-the-art
inverted index-based solutions and further outperforms the winning
(graph-based) submissions to the BigANN Challenge by a significant margin.
\end{abstract}



\maketitle

\section{Introduction}
\label{sec:introduction}

Neural Information Retrieval (NIR) has gained increasing popularity since the introduction of
pre-trained Large Language Models (LLMs)~\cite{DBLP:series/synthesis/2021LinNY}. NIR models learn a vector
representation of short pieces of text, known as an \emph{embedding}, that captures the contextual semantics of the input,
thereby enabling more effective matching of queries to documents and, thus, first-stage retrieval~\cite{INR-071}.

One major focus in NIR is what we call \emph{learned sparse retrieval}
(LSR)~\cite{epic,splade-sigir2021,formal2021splade,formal2022splade,lassance2022efficient-splade}.
LSR repurposes an LLM to encode an input into \emph{sparse} embeddings,
a vector in an inner product space where each dimension corresponds with a term
in the model's vocabulary. When a coordinate is nonzero in an embedding, that indicates
that the corresponding term is semantically relevant to the input.
Similarity between embeddings is typically determined by inner product, so that
retrieval given a query becomes the problem known as Maximum Inner Product Search (MIPS):
Finding the top-$k$ vectors that maximize inner product with a query vector.

LSR is attractive for three reasons.
First, LSR models are competitive with \emph{dense retrieval}
models that encode text into dense
vectors~\cite{DBLP:series/synthesis/2021LinNY,karpukhin-etal-2020-dense,xiong2021approximate,reimers-2019-sentence-bert,santhanam-etal-2022-colbertv2,colbert2020khattab,10.1007/978-3-031-56060-6_1}. Importantly, evidence suggests that
some LSR models generalize better to out-of-domain datasets~\cite{bruch2023fusion,lassance2022efficient-splade}.

Second, because of the one-to-one mapping between dimensions and vocabulary terms,
sparse embeddings are \emph{interpretable} by design. A user can easily understand
the embedding space, explain retrieval results, and debug relevance issues.
Such properties may be of interest in medical and security applications, for example.

The final reason for their popularity is that sparse embeddings
retain many of the benefits of classical lexical models such as BM25~\cite{bm25original}
while addressing one of their major weaknesses. That is because, sparse embeddings
can, at least in theory, be indexed and retrieved using the all-too-familiar inverted index-based
machinery~\cite{tonellotto2018survey}, while at the same time, remedying the \emph{vocabulary mismatch} problem
due to the incorporation of contextual signals.

Their performance, interpretability, and similarity to lexical models
make LSR an important area of research. Efforts in this space include improving the
effectiveness of sparse embeddings~\cite{formal2022splade,formal2021splade}
and the efficiency of sparse retrieval
algorithms~\cite{bruch2023sinnamon,bruch2023bridging,formal2023tois-splade,10.1145/3576922,mallia2022guided-traversal}.

The latter category is justified because, despite the apparent compatibility of sparse embeddings with
inverted indexes, efficient retrieval remains a challenge. 
That is so because the weights learned by LSR models exhibit statistical properties that do not conform
to the assumptions under which popular inverted index-based retrieval algorithms operate~\cite{bruch2023sinnamon,mackenzie2021wacky,crane2017wsdm}.
For example, algorithms such as WAND~\cite{broder2003wand} and MaxScore~\cite{maxscore}, that are
designed for term frequency-based lexical models, function far better than their worst-case complexity would suggest,
\emph{if} queries are short and term frequencies follow a Zipfian distribution.
In LSR, queries are often longer and, crucially,
frequencies are no longer Zipfian~\cite{bruch2023sinnamon}.
That deviation from assumptions often translates to increased per-query latency.

Overcoming these limitations requires either forcing LSR models to produce the ``right'' distribution,
or designing retrieval algorithms that have fewer restrictive assumptions.
As an example of the first direction, Efficient \splade{}~\cite{lassance2022efficient-splade} applies $L_1$ regularization and 
uses dedicated query and document encoders to make queries shorter.
As another,~\cite{lassance2023static-pruning} statically prunes documents
(or inverted lists) to produce embeddings that approximately maintain semantics
but with statistics that are more friendly to dynamic pruning algorithms.

Works in the second direction~\cite{bruch2023sinnamon,bruch2023bridging} take a leaf out
of the Approximate Nearest Neighbor (ANN) literature~\cite{bruch2024foundations}:
Algorithms that produce \emph{approximate}, as opposed to \emph{exact},
top-$k$ sets. This relaxation makes it easier to trade off accuracy
for large gains in efficiency.

Approximate retrieval offers great potential and serves as a bridge between
dense and sparse retrieval~\cite{bruch2023bridging}. So appealing is this paradigm
that the 2023 BigANN Challenge\footnote{\url{https://big-ann-benchmarks.com/neurips23.html}}
at NeurIPS dedicated a track to learned sparse embeddings.
Submissions were evaluated on the \splade{}~\cite{formal2023tois-splade}
embeddings of the \msmarco~\cite{nguyen2016msmarco} Passage dataset,
and were ranked by the highest throughput past $90\%$ accuracy (i.e., recall with respect to exact search).
The results were intriguing: the top two
submissions were graph-based ANN methods designed for dense vectors,
while other approaches, including an optimized approximate inverted index-based design struggled.

Inspired by BigANN, we present a novel ANN algorithm that we call \our
(\textbf{S}pilled Clust\textbf{e}ring of \textbf{I}nverted Lists with \textbf{S}ummaries
for \textbf{M}aximum \textbf{I}nner Produ\textbf{c}t Search)
and that admits effective and efficient retrieval over learned sparse embeddings.
Pleasantly, our design uses in a new way two familiar data structures: the inverted and the forward index.
In particular, we extend the inverted index by introducing a novel organization of inverted lists into geometrically-cohesive blocks.
Each block is equipped with a ``sketch,'' serving as a \emph{summary} of the vectors contained in it.
The summaries allow us to skip over a large number of blocks during retrieval and save substantial compute.
When a summary indicates that a block must be examined, we use the forward index
to retrieve exact embeddings of its documents and compute inner products.

We evaluate \our against strong baselines, including the top (open-source) submissions to the BigANN Challenge.
We additionally include classic inverted index-based retrieval and impact-sorted indexes as reference points for completeness.
Experimental results show average per-query latency in \textbf{microsecond territory} on various sparse embeddings of
\msmarco~\cite{nguyen2016msmarco}.
Impressively, \our \textbf{outperforms the graph-based winning solutions of
the BigANN Challenge by a factor of at least 3.4 at 95\% accuracy on \splade and 12 on Efficient \splade},
with the margin widening substantially as accuracy increases. Other baselines, including state-of-the-art
inverted index-based algorithms, are \textbf{consistently one to two orders of magnitude slower than \our}.

In summary, we make the following contributions in this work:
\begin{itemize}[leftmargin=*]
    \item We study an empirical property of learned sparse embeddings
    that we call the ``concentration of importance'';
    \item We present \our, a novel ANN algorithm for retrieval over learned sparse vectors
    that is based on a geometrical organization of the inverted index, and leverages the
    concentration of importance;
    \item We report, through extensive experiments, remarkable gains in query latency
    in exchange for a negligible loss in \emph{retrieval} accuracy, outperforming several state-of-the-art
    baselines, including the winning submissions to the 2023 BigANN Challenge; and,
    \item We given an in-depth analysis of \our in an ablation study.
\end{itemize}

\section{Related Work}
\label{sec:related}
This section reviews notable related research.
We summarize the thread of work on learned sparse embeddings,
then discuss methods that approach the problem of retrieval
over such vector collections.

\subsection{Learned Sparse Representations}
Learned sparse representations were investigated~\cite{zamani2018sparse}
even before the emergence of pre-trained LLMs.
But the rise of LLMs supercharged this research and led to a flurry of activity on the
topic~\cite{dai2019contextaware,10.1145/3366423.3380258,10.1145/3397271.3401204,epic,zhao-etal-2021-sparta,sparterm,formal2021splade,formal2023tois-splade,unicoil}.
First attempts at this include DeepCT and HDCT by Dai and Callan~\cite{dai2019contextaware,10.1145/3366423.3380258,10.1145/3397271.3401204}.

DeepCT used the Transformer~\cite{vaswani2017attention} encoder of BERT~\cite{devlin2019bert}
to extract contextual features of a word into an embedding,
which can be viewed as a feature vector that characterizes the term's syntactic
and semantic role in a given context.
DeepCT linearly combines a term's contextualized embedding and summarizes it as a term \emph{weight}
for terms that are present in a document.
Because the vocabulary associated with a document remains the same, it does not
address the vocabulary mismatch problem.

One way to address vocabulary mismatch is to use a generative model,
such as doc2query~\cite{nogueira2019document} or docT5query~\cite{nogueira2019doc2query},
to expand documents with relevant terms \emph{and} boost existing terms
by repeating them in the document, implicitly performing term re-weighting.
In fact, \unicoil~\cite{unicoil,ma2022document} expands its input with DocT5Query~\cite{nogueira2019doc2query}
before learning and producing a sparse representation.

Formal \etal build on SparTerm~\cite{sparterm} and propose \splade{}~\cite{splade-sigir2021}.
Their construction introduces sparsity-inducing regularization and a
log-saturation effect on term weights, so that the sparse representations learned by \splade{}
are typically relatively sparser. Interestingly, \splade{} showed competitive results with respect to
state-of-the-art dense and sparse methods~\cite{splade-sigir2021}.

In a later work, Formal \etal make adjustments to \splade{}'s
pooling and expansion mechanisms, and introduce distillation into its training.
This second version, called \splade{} v2, reached state-of-the-art results on the
\msmarco{}~\cite{nguyen2016msmarco} passage ranking task as well as the \beir{}~\cite{thakur2021beir}
zero-shot evaluation benchmark~\cite{formal2021splade}.
The \splade{} model has undergone many other rounds of improvements which have been documented
in the latest work by the same authors~\cite{formal2023tois-splade}.
Among these, one notable extension is the Efficient \splade{} which, as we already noted,
attempts to make the learned embeddings more friendly to inverted index-based algorithms.

\subsection{Retrieval Algorithms}

The Information Retrieval literature offers a wide array
of algorithms tailored to retrieval on text collections~\cite{tonellotto2018survey}.
They are often \emph{exact} and scale easily to massive datasets.
MaxScore~\cite{maxscore} and WAND~\cite{broder2003wand}, and
subsequent improvements~\cite{ding2011bmwand,topk_bmindexes,mallia2019faster-blockmaxwand,mallia2017blockmaxwand_variableBlocks},
are examples that, essentially, solve the MIPS problem
over ``bag-of-words'' representations of text, such as
BM25~\cite{bm25original} or TF-IDF~\cite{salton1988term}.

These algorithms operate on an inverted index, augmented with additional data
to speed up query processing.
One that features prominently is the maximum attainable
partial inner product---an upper-bound.
This enables the possibility of navigating the inverted lists, one document
at a time, and deciding quickly if a document may belong to the result set.
Effectively, such algorithms (safely) \emph{prune} the parts of the index
that cannot be in the top-$k$ set. That is why they are
often referred to as \emph{dynamic pruning} techniques.

Although efficient in practice, dynamic pruning methods are designed specifically for text collections.
Importantly, they ground their performance on
several pivotal assumptions: non-negativity, higher sparsity rate for queries,
and a Zipfian shape of the length of inverted lists.
These assumptions are valid for TF-IDF or BM25,
which is the reason why dynamic pruning works well and the worst-case
time complexity of MIPS is seldom encountered in practice.

These assumptions do not necessarily hold for collections of learned sparse representations, however.
Learned vectors may be real-valued, with a sparsity rate that is closer to uniform across
dimensions~\cite{bruch2023sinnamon,mackenzie2021wacky}.
Mackenzie \etal~\cite{10.1145/3576922} find that learned sparse embeddings reduce the odds of pruning
or early-termination in the document-at-a-time (DaaT) and Score-at-a-Time (SaaT) paradigms.

The most similar work to ours is~\cite{bruch2023bridging}.
The authors investigate if \emph{approximate} MIPS algorithms for \emph{dense} vectors
port over to \emph{sparse} vectors.
They focus on \emph{inverted file} (IVF) where vectors are partitioned into clusters
during indexing, with only a fraction of clusters scanned during retrieval.
They show that IVF serves as an efficient solution for sparse MIPS.
Interestingly, the authors cast IVF as dynamic pruning and
turn that insight into a novel organization of the inverted index for
approximate MIPS for general sparse vectors.
Our index structure can be viewed as an extension of theirs.

Finally, we briefly describe another ANN algorithm over dense vectors:
HNSW~\cite{hnsw2020}, a graph-based algorithm that constructs a graph
where each document is a node and two nodes are connected
if they are deemed ``similar.'' Similarity is based on Euclidean distance,
but~\cite{ip-nsw18} shows inner product results in a structure that is
capable of solving MIPS rather quickly and accurately.
As we learn in the presentation of our empirical analysis,
algorithms that adapt IP-HNSW~\cite{ip-nsw18} to sparse vectors work remarkably well.

\section{Definitions and Notation}
\label{sec:definition}
Suppose we have a collection $\mathcal{X} \subset \mathbb{R}_{+}^{d}$ of
nonnegative \emph{sparse} vectors. If $x \in \mathcal{X}$,
then $x$ is a $d$-dimensional vector where the vast majority of its coordinates are $0$
and the rest are real positive values.
We use superscript to enumerate a collection:
$x^{(j)}$ is the $j$-th vector in $\mathcal{X}$.

We use lower-case letters (e.g., $x$) to denote a vector,
call $1 \leq i \leq d$ its \emph{coordinate},
and write $x_i$ for its $i$-th \emph{value}.
Together, we refer to a coordinate and value pair as an \emph{entry},
and say an entry is non-zero if it has a non-zero value.
A sparse vector can be identified as a set of non-zero entries:
$\{ (i, x_i) \;|\; x_i \neq 0 \}$.

Sparse MIPS aims to solve the following problem
to find, from $\mathcal{X}$, the set $\mathcal{S}$ of top $k$ vectors
whose inner product with the query vector $q \in \mathbb{R}^{d}$ is maximal:
\begin{equation}
    \mathcal{S} = \argmax^{(k)}_{x \in \mathcal{X}} \; \langle q, x \rangle.
    \label{equation:mips}
\end{equation}

Let us define a few concepts that we frequently refer to.
The $L_p$ norm of a vector denoted by $\lVert \cdot \rVert_p$ is defined as
$\lVert x\rVert_p = (\sum_i \lvert x_i\rvert^p)^{1/p}$. We call the $L_p$ norm
of a vector its $L_p$ \emph{mass}. Additionally:

\begin{definition}[$\alpha$-mass subvector]
    Consider a vector $x$ and a permutation $\pi$
    that sorts the entries of $x$ by their absolute value:
    $\lvert x_{\pi_i} \rvert \geq \lvert x_{\pi_{i+1}} \rvert$.
    For a constant $\alpha \in [0, 1]$, denote by $1 \leq j \leq d$
    the smallest integer such that:
    \begin{equation*}
        \sum_{i=1}^j \lvert x_{\pi_i} \rvert \leq \alpha \lVert x \rVert_1.
    \end{equation*}
    We call $\tilde{x}$ made up of $\{ (\pi_i, x_{\pi_i}) \}_{i=1}^j$,
    the \emph{$\alpha$-mass subvector} of $x$.
    Clearly, $\lVert \tilde{x} \rVert_1 \leq \alpha \lVert x \rVert_1$.
\end{definition}

\section{Concentration of Importance}
\label{subsec:property}

Recently, Daliri \etal~\cite{daliri2023sampling} presented a sketching algorithm
for sparse vectors that rest on the following simple principle: Coordinates
that contribute more heavily to the $L_2$ norm of a vector, weigh more
significantly on the inner product between vectors. Using that intuition,
they report that if we were to drop the non-zero coordinates of a sparse vector
with a probability proportional to its contribution to the $L_2$ mass, we can reduce the size of a collection while approximately maintaining inner products between vectors.

Inspired by~\cite{daliri2023sampling}, we examined two state-of-the-art LSR techniques:
\splade{}~\cite{formal2022splade} and Efficient \splade{}~\cite{lassance2022efficient-splade}.
Our analysis reveals a parallel property, which we call the ``concentration of
importance.''
In particular, we observe that the LSR techniques place
a disproportionate amount of the total $L_1$ mass of a vector
on just a small subset of the coordinates.

\begin{figure}[t]
\centering
\includegraphics[width=0.8\columnwidth]{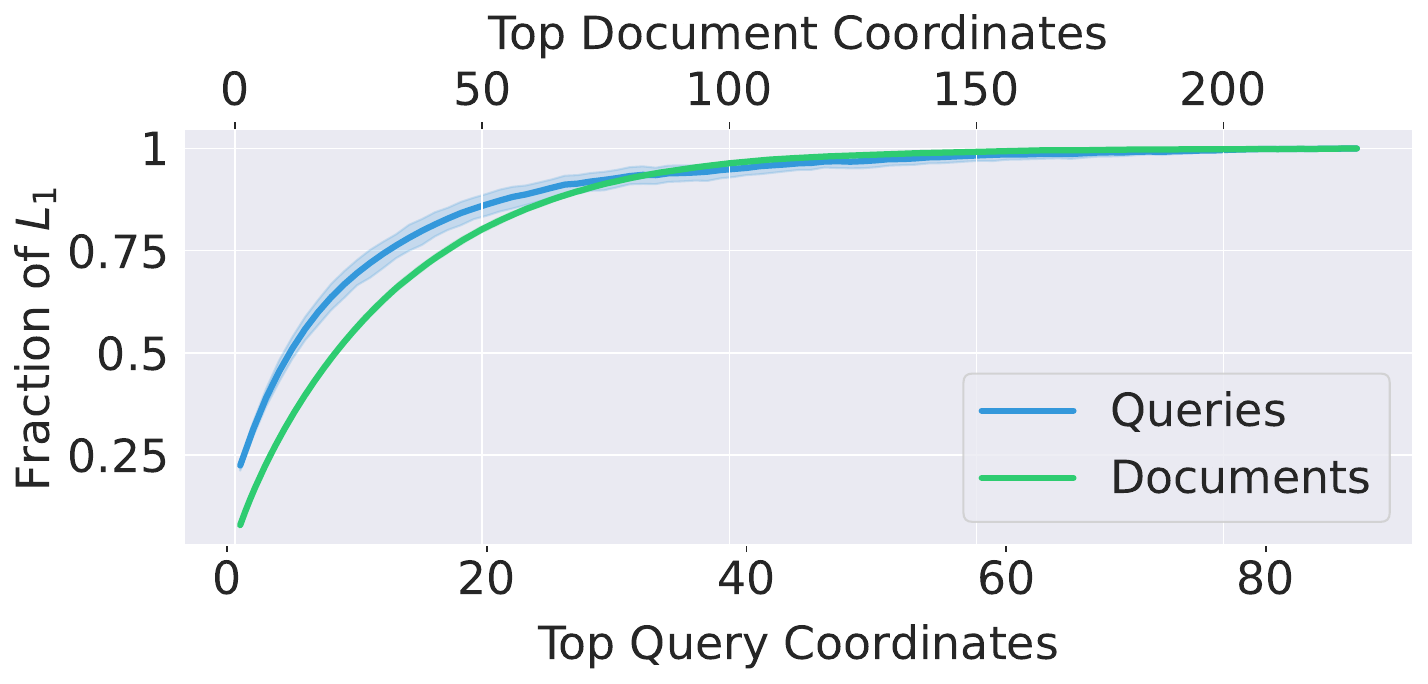}
\caption{Fraction of $L_1$ mass preserved by keeping only the top
non-zero entries with the largest absolute value.
\label{fig:energy}}
\end{figure}

Let us demonstrate this phenomenon on the \msmarco Passage dataset~\cite{nguyen2016msmarco}
with the \splade{} embeddings.\footnote{The \texttt{cocondenser-ensembledistill} checkpoint was obtained from \url{https://huggingface.co/naver/splade-cocondenser-ensembledistil}.}
We take every vector, sort its entries by value,
and measure the fraction of the $L_1$ mass preserved by considering a given number of top entries.
For queries, the top $10$ entries yield $0.75$-mass subvectors.
For documents, the top $50$ (about $30$\% of non-zero entries),
give $0.75$-mass subvectors.
We illustrated our measurements in Figure~\ref{fig:energy}.

\begin{figure}[t]
\centering
\includegraphics[width=0.8\columnwidth]{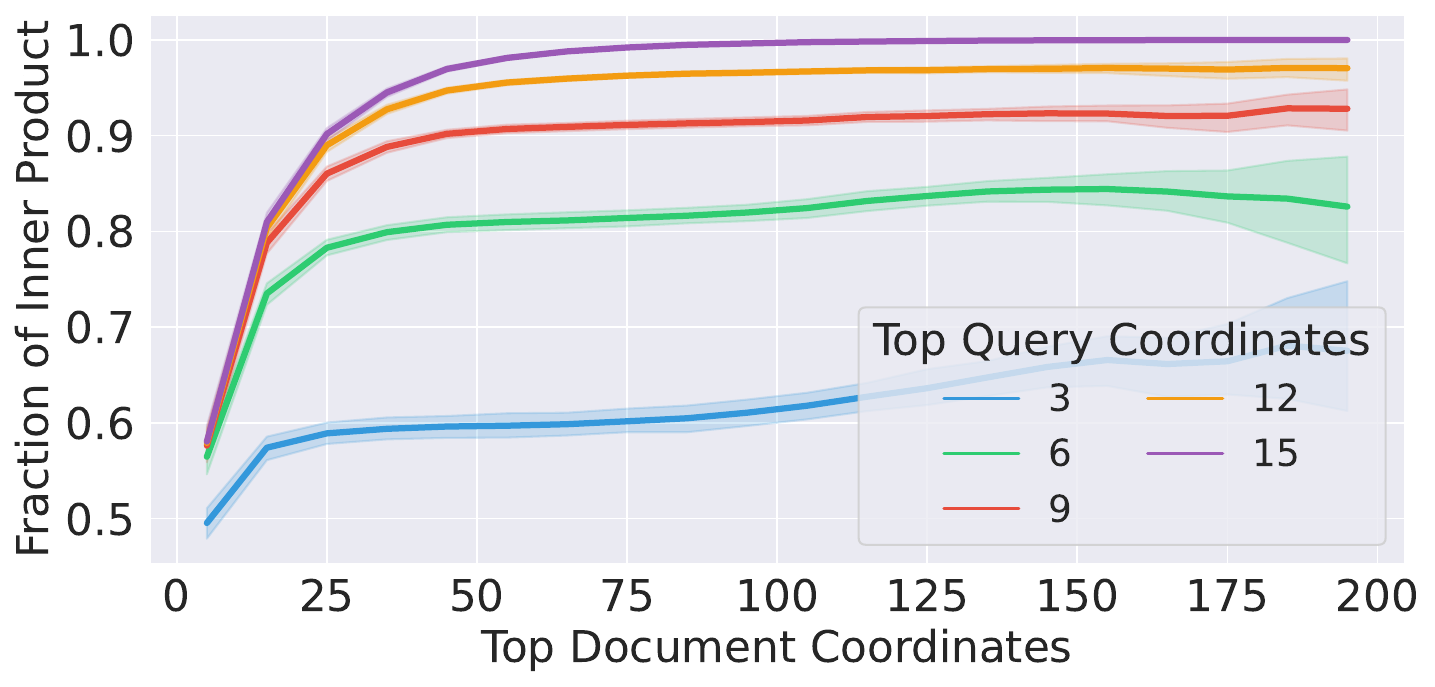}
\caption{Fraction of inner product (with $95\%$
confidence intervals) preserved by inner product between
the top query and document coordinates with the largest absolute value.
\label{fig:energy2}}
\end{figure}

These results bring us naturally to our next question:
What are the ramifications of the concentration of importance for inner product
between queries and documents?
One way to study that is as follows: We take the top-$10$ document vectors for each query,
prune each document vector by keeping a fraction of its non-zero entries with the largest value.
We do the same for query vectors.
We then compute the inner product between the trimmed-down queries and documents
and report the results in Figure~\ref{fig:energy2}.

The figure shows that, if we took the top $10\%$ of the most ``important'' coordinates
from queries ($9$) and documents ($20$), we preserve, on average,
$85\%$ of the full inner product.
Keeping $12$ query and $25$ document coordinates bumps that up to $90\%$.

Our results confirm that LSR tends to concentrate importance on a few coordinates.
Furthermore, a partial inner product between the largest entries (by absolute value)
approximates the full inner product with arbitrary accuracy.
As we will see shortly, this property, which is in agreement with~\cite{daliri2023sampling},
can help speed up query processing and reduce space consumption rather substantially.

\section{Proposed Algorithm}
\label{sec:methodology}

\begin{figure*}[t]
\centering
\includegraphics[width=1.7\columnwidth]{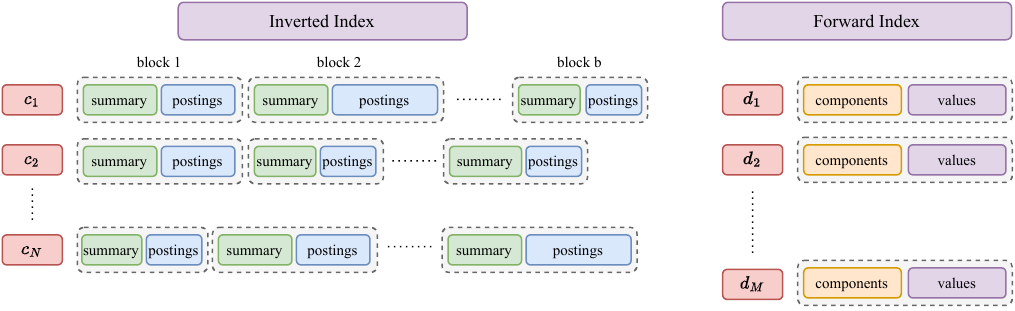}
\caption{The design of \our. Inverted lists are independently partitioned into geometrically-cohesive blocks.
Each block is a set of document identifiers with a summary vector.
The inner product of a query with the summary
approximates the inner product attainable with the documents in that block.
The forward index stores the complete vectors (including values).}
\label{fig:inverted}
\end{figure*}

We now introduce \our, a novel ANN algorithm that allows effective and efficient
approximate retrieval over learned sparse representations. The design of \our 
uses two important and familiar data structures: the inverted index and the forward index.
In an nutshell, we use a forward index for inner product computation,
and an inverted index to pinpoint the subset of documents that must be evaluated.
Figure \ref{fig:inverted} gives an overview of the overall design.

\our is novel in the following ways.
First, it uses an organization of the inverted index that blends together \emph{static}
and \emph{dynamic} pruning to significantly reduce the number of documents that must be evaluated
during retrieval. Second, it partitions inverted lists into geometrically-cohesive blocks to facilitate
efficient skipping of blocks. Finally, we attach a \emph{summary} to each block,
whose inner product with a query approximates---albeit not necessarily in an unbiased
manner---the inner product of the query with documents contained in the block.

\subsection{Static Pruning}
\our heavily relies on the concentration of importance property discussed in Section~\ref{subsec:property}.
The property shows that a small subset of the most important coordinates of the sparse embedding
of a query and document vector can be used to effectively approximate their inner product.
We incorporate this result in \our during the construction of the inverted index
through \emph{static pruning}.

Concretely, for coordinate $i$,
we build its inverted list by gathering all $x \in \mathcal{X}$ whose $x_i \neq 0$.
We then sort the inverted list by $x_i$'s value in decreasing order (breaking ties arbitrarily),
so that the document whose $i$-th coordinate has the largest value appears at the beginning of the list.
We then prune the inverted list by keeping at most the first $\lambda$ entries for a fixed
$\lambda$---our first hyper-parameter.
We denote the resulting inverted list for coordinate $i$ by $\mathcal{I}_i$.

\subsection{Blocking of Inverted Lists}
\label{sec:methodology:blocking}
\our also introduces a novel blocking strategy on inverted lists.
It partitions each inverted list into $\beta$ small blocks---our second hyper-parameter.
The rationale behind a blocked organization of an inverted list is to group together
documents that are \emph{similar} in terms of their local representations,
so as to facilitate a \emph{dynamic pruning} strategy, to be described shortly.

We defer the determination of similarity to a clustering algorithm.
In other words, the documents whose ids are present in an inverted list are given as input to a clustering
algorithm, which subsequently partitions them into $\beta$ clusters.
Each cluster is then turned into one block, consisting of the id of documents
whose vectors belong to the same cluster.
Conceptually, each block is ``atomic'' in the following sense: if the dynamic pruning algorithm decides
we must visit a block, \emph{all} the documents in that block are fully evaluated.

We note that any geometrical (supervised or unsupervised) clustering algorithm may be readily used.
We use a shallow variant~\cite{chierichetti2007clusterPruning} of K-Means as follows.
Given a set of vectors $\mathcal{S}$, we uniformly-randomly sample $\beta$ vectors, $\{ \mu^{(j)} \}_{j=1}^\beta$,
from $\mathcal{S}$, and use them as cluster representatives.
For each $x \in \mathcal{S}$, we find $j^\ast = \argmax_j \langle x, \mu^{(j)} \rangle$,
and assign $x$ to the $j^\ast$-th cluster.

\subsection{Per-block Summary Vectors}
So far we have described how we statically prune inverted lists to the top $\lambda$ entries
and then partition them into $\beta$ blocks using a clustering algorithm. 
We now describe how this structure can be used as a basis for a novel dynamic pruning method.

We need an efficient way to determine if a block should be evaluated.
To that end, \our leverages the concept of a \emph{summary} vector:
a $d$-dimensional vector that ``represents'' the documents in a block.
The summary vectors are stored in the inverted index, one per block,
and are meant to serve as an efficient way to compute
a good-enough approximation of the inner product between a query and the documents within the block.

One realization of this idea is to upper-bound the full inner product
attainable by documents in a block.
In other words, the $i$-th coordinate of the summary vector of a block would contain
the maximum $x_i$ among the documents in that block.
This construction can be best described as
a vectorization of the upper-bound \emph{scalars} in blocked variants of WAND~\cite{ding2011bmwand}.

More precisely, our summary function $\phi: 2^\mathcal{X} \rightarrow \mathbb{R}^d$
takes a block $B$ from the universe of all blocks $2^\mathcal{X}$,
and produces a vector whose $i$-th coordinate is simply:
\begin{equation}
    \label{equation:summary}
    \phi(B)_i = \max_{x \in B} x_i.
\end{equation}
This summary is \emph{conservative}:
its inner product with the query is no less than the inner product
between the query and any of its document: $\langle q, \phi(B) \rangle \geq \langle q, x \rangle$ 
for all $x \in B$ and an arbitrary query $q$.

The caveat, however, is that the number of non-zero entries in summary vectors
grows quickly with the block size. 
That is the source of two potential issues: 1) the space required to store summaries
increases; and 2) as inner product computation takes time proportional
to the number of non-zero entries, the time required to evaluate 
a block could become unacceptably high.

We may address that caveat by applying pruning and quantization,
with the understanding that any such method may take away
the conservatism of the summary.
As we will empirically show, there are many pruning or quantization candidates
to choose from.

In particular, we use the following technique that
builds on the concentration of importance property: We prune $\phi(B)$,
obtained from Equation~(\ref{equation:summary}), by keeping only its $\alpha$-mass subvector.
That, $\alpha$, is our third and last indexing hyper-parameter.

We further reduce the size of summaries by applying scalar quantization.
With the goal of reserving a single byte for each value, we 
subtract the minimum value $m$ from each summary entry,
and divide the resulting range into $256$ sub-intervals of equal size.
A value in the summary is replaced with the index of the sub-interval it maps to.
To reconstruct a value approximately, we multiply the id of its sub-interval
by the size of the sub-intervals, then add $m$.

\subsection{Forward Index}
\our blends together two data structures.
The first is an inverted index that tells us which documents to examine.
To make it practical, we apply approximations that allow us
to gain efficiency with a possible loss in accuracy.
A forward index, which is simply a look-up table that stores the
exact document vectors, helps correct those errors and offers a way to compute the
exact inner products between a query and the documents within a block, whenever
that block is deemed a good candidate for evaluation.

We must note that, documents belonging to the same block are not necessarily stored consecutively
in the forward index. This is simply infeasible because the same document may belong to
different inverted lists and, thus, to different blocks. Because of this layout,
computing the inner products may incur many cache misses, which are detrimental to query latency.
In our implementation, we extensively use prefetching instructions to mitigate this effect.

\subsection{Recap}
We summarize the discussion above in Algorithm \ref{algorithm:indexing}.
When indexing a collection $\mathcal{X} \subset \mathbb{R}^d$,
for every coordinate $i \in \{ 1, \dots, d\}$, we form its inverted list,
recording only the document identifiers (Line~\ref{algorithm:indexing:inverted-list}).
We then sort the list in decreasing order of values (Line~\ref{algorithm:indexing:sort}), and
apply static pruning by keeping, for each inverted list, the $\lambda$ elements with
the largest value (Line~\ref{algorithm:indexing:static-pruning}).
We then apply clustering to the inverted list to derive at most $\beta$ blocks
(Line~\ref{algorithm:indexing:clustering}).
Once documents are assigned to the blocks, we then build the block summary
using the procedure described earlier (Line~\ref{algorithm:indexing:summary}).

Algorithm \ref{algorithm:retrieval} shows the query processing logic in \our.
We use the concentration of importance property to (a) select a subset
of the query coordinates $q_\cut$ (Line~\ref{algorithm:retrieval:q-cut}),
and (b) define a novel dynamic pruning strategy
(Lines~\ref{algorithm:retrieval:summary-ip}--\ref{algorithm:retrieval:skip})
that allows to skip blocks in the inverted lists of the coordinates in $q_\cut$.

\begin{algorithm}[t]
\SetAlgoLined
{\bf Input: }{
Collection $\mathcal{X}$ of sparse vectors in $\mathbb{R}^{d}$;
$\lambda$: Maximum length of each inverted list;
$\beta$: Maximum number of blocks per inverted list;
$\alpha$: Fraction of the overall importance preserved by each summary.\\}
\KwResult{\our index.}
\begin{algorithmic}[1]
    \FOR{$i \in \{ 1, \ldots, d\}$}
        \STATE $\mathcal{S} \leftarrow \{ j \;|\; x^{(j)}_i \neq 0,\; x^{(j)} \in \mathcal{X} \}$
            \label{algorithm:indexing:inverted-list}
        \STATE \textsc{Sort} $\mathcal{S}$ in decreasing order by $x_i$ for all $x \in \mathcal{S}$
            \label{algorithm:indexing:sort}
        \STATE $\mathcal{I}_i \leftarrow \{ \mathcal{S}_{i,1}, \mathcal{S}_{i,2}, \ldots, \mathcal{S}_{i,\lambda} \}$
            \label{algorithm:indexing:static-pruning}
        \STATE \textsc{Cluster} $\mathcal{I}_i$ into $\beta$ partitions, $\{ B_{i, j} \}_{j=1}^\beta$
            \label{algorithm:indexing:clustering}
        \FOR{$1 \leq j \leq \beta$}
            \STATE $S_{i, j} \leftarrow \alpha$-mass subvector of $\phi(B_{i, j})$
            \COMMENT{Equation~(\ref{equation:summary})}
                \label{algorithm:indexing:summary}
        \ENDFOR
    \ENDFOR
    \RETURN $\mathcal{I}_i$, $\{ S_{i, j} \} \; \forall i, j$
 \end{algorithmic}
 \caption{Indexing with \our.}
\label{algorithm:indexing}
\end{algorithm}

\our adopts a coordinate-at-a-time traversal (Line~\ref{algorithm:retrieval:traversal}) of the inverted index.
For each coordinate $i \in q_\cut$, it evaluates the blocks using their summary.
The documents within a block are evaluated further if the approximation with
the summary is greater than a fraction of the minimum inner product in the Min-\heap.
That means that, the forward index retrieves the complete document vector
in the selected block and computes inner products.
A document whose inner product is greater than the minimum score in
the Min-\heap{} is inserted into the heap.
Note that, Algorithm~\ref{algorithm:retrieval} takes two hyper-parameters: an integer $\cut$,
and $\heapfactor \in (0, 1)$.

\begin{algorithm}[t]
\SetAlgoLined
{\bf Input: }{$q$: query;
$k$: number of results;
\cut: number of largest query entries considered;
{\heapfactor}: a correction factor to rescale the summary inner product;
$\mathcal{I}_i$'s and $S_{i, j}$'s: inverted lists and summaries obtained from
Algorithm~\ref{algorithm:indexing}.}\\
\KwResult{ A \heap\ with the top-$k$ documents.}
\begin{algorithmic}[1]
\STATE $q_\cut \leftarrow$ the top \cut\ entries of $q$ with the largest value
    \label{algorithm:retrieval:q-cut}
\STATE \heap $\leftarrow \emptyset$
\FOR{$i \in q_\cut$}
    \label{algorithm:retrieval:traversal}
    \FOR{$B_j \in \mathcal{I}_i$}
        \STATE $r \leftarrow \langle q, S_{i, j} \rangle$
            \label{algorithm:retrieval:summary-ip}
        \IF{$\heap.{\sf len()} = k$ and  $r < \frac{\heap.{\sf min()}}{\heapfactor}$ }
            \STATE {\bf continue} \COMMENT{Skip the block}
            \label{algorithm:retrieval:skip}
        \ENDIF
        \FOR{$d \in B_j$}
            \STATE $p = \langle q, {\sf ForwardIndex}[d] \rangle$
            \IF{$\heap.{\sf len()} < k$ or $p > \heap.{\sf min()}$ }
                \STATE \heap.{\sf insert}$(p, d)$
            \ENDIF
            \IF{$\heap.{\sf len()} = k+1$}
                \STATE \heap.{\sf pop\_min()}
            \ENDIF
        \ENDFOR
    \ENDFOR
\ENDFOR
\RETURN \heap
\end{algorithmic}
\caption{Query processing with \our.\label{algorithm:retrieval}}
\end{algorithm}

\section{Generalized Architecture}
\label{sec:architecture}

What we presented in Section~\ref{sec:methodology} is an instance of a more general algorithm.
Conceptually, \our{} can be viewed as the application of the following logical functions
to a collection of sparse vectors.

\vspace{1mm}
\noindent \textbf{Clustering with Spillage}. 
We group together documents that share a non-zero coordinate (as inverted lists),
then partition them into blocks. This is an instance of
\emph{clustering with spillage}, where an item may belong to multiple clusters.
The inverted index as \emph{coarse} clustering is efficient for sparse vectors, though
other algorithms that allow spillage may very well suit other distributions.

\vspace{1mm}
\noindent \textbf{Sketching}. We summarize clusters by taking the maximum
of each coordinate.
While we use the upper-bound vector to obtain a conservative estimate,
a more general design admits other types of summaries such as centroids, medoids
or any other sketch~\cite{woodruff2014sketching}.

\vspace{1mm}
\noindent \textbf{Compression}. We used pruning and quantization to
reduce the total size of summaries by paying particular attention to the $L_1$ mass.
In theory, however, any number of other compression schemes may be utilized,
such as~\cite{bruch2023sinnamon,daliri2023sampling}.

\vspace{1mm}
\noindent \textbf{Routing}. We identify the subset of clusters that must be fully evaluated by
sequentially scanning summaries and comparing their inner product with the minimum score so far.
Routing a query to the right cluster, however, need not follow that paradigm strictly.
We may consider all summaries at once and decide which clusters to probe
in one go---a process akin to the ``IVF'' approach to ANN~\cite{pq}.

\section{Experiments}
\label{sec:experiments}

We now evaluate \our{} experimentally. Specifically, we are interested in investigating the performance of \our in the following ways:
(a) its accuracy, latency, space usage, and indexing time against existing solutions, and
(b) an ablation study of the impact of the different components of \our on performance.

In what follows, we unpack these questions through
an empirical evaluation on two public datasets. We note that, due to space constraints,
we excluded many combinations of datasets and LSR models (e.g., \unicoil embeddings of \nq)
from the presentation of our results. However, the reported trends hold consistently.

\subsection{Setup}

\noindent \textbf{Datasets}.
We experiment on two publicly-available datasets: \msmarco{} v1
Passage~\cite{nguyen2016msmarco} and Natural Questions (\nq{}) from \beir~\cite{thakur2021beir}.
\msmarco{} is a collection of $8.8$M passages in English. In our evaluation,
we use the smaller ``dev'' set of queries for retrieval, which includes $6{,}980$ questions.
\nq{} is a collection of $2.68$M questions in English. We use it in combination with its
``test'' set of $7{,}842$ queries.

\vspace{1mm}
\noindent \textbf{Learned Sparse Representations}.
We evaluate \our with embeddings generated by three LSR models:
\begin{itemize}[leftmargin=*]
\item \splade~\cite{formal2022splade}. Each non-zero entry is the importance weight of a term in the BERT~\cite{devlin2019bert}
WordPiece~\cite{wordpiece} vocabulary consisting of $30$,$000$ terms.
We use the \texttt{cocondenser-ensembledistil}\footnote{Checkpoint at \url{https://huggingface.co/naver/splade-cocondenser-ensembledistil}}
version of \splade that yields MRR@10 of $38$.$3$ on the \msmarco dev set.
The number of non-zero entries in documents (queries) is, on average,
$119$ ($43$) for \msmarco and $153$ ($51$) for \nq.

\item Efficient \splade~\cite{lassance2022efficient-splade}. Similar to \splade, 
but there are $181$ ($5.9$) non-zero entries in \msmarco documents (queries).
We use the \texttt{efficient-splade-V-large}\footnote{Checkpoints at \url{https://huggingface.co/naver/efficient-splade-V-large-doc} and \url{https://huggingface.co/naver/efficient-splade-V-large-query}.}
version, yielding MRR@10 of $38$.$8$ on the \msmarco dev set.
We refer to this model as \esplade{}.

\item \unicoil~\cite{unicoil,ma2022document}. Expands passages with relevant terms generated by
DocT5Query~\cite{nogueira2019doc2query}. \unicoil achieves MRR@10 of $35$.$2$ on the \msmarco dev set.
There are, on average, $68$ ($6$) non-zero entries in \msmarco documents (queries).
\end{itemize}

It is worth highlighting that these embedding models belong to different families.
\splade and \esplade perform expansion for both queries and documents.
On the other hand, \unicoil only performs document expansion and does so using a generative model.

We generate the \splade and \esplade embeddings using the original code published on GitHub.\footnote{\url{https://github.com/naver/splade}}
\unicoil embeddings are based on the original implementation on GitHub.\footnote{\url{https://github.com/castorini/pyserini/blob/master/docs/experiments-unicoil.md}}
After generating the embeddings,
we replicate the performance in terms of MRR@10 on the \msmarco dev set to confirm that our replication achieves the same performance presented in the original papers.

\begin{table*}[t]
	\centering
    \adjustbox{max width=\textwidth}{%
   	\begin{tabular}{lr@{\hspace{.75\tabcolsep}}rr@{\hspace{.75\tabcolsep}}rr@{\hspace{.75\tabcolsep}}rr@{\hspace{.75\tabcolsep}}rr@{\hspace{.75\tabcolsep}}rr@{\hspace{.75\tabcolsep}}rr@{\hspace{.75\tabcolsep}}rr@{\hspace{.75\tabcolsep}}r} 

        \toprule
        \multicolumn{17}{c}{\splade on \msmarco} \\
        \midrule
        Accuracy (\%) & \multicolumn{2}{c}{90} & \multicolumn{2}{c}{91} & \multicolumn{2}{c}{92} & \multicolumn{2}{c}{93} & \multicolumn{2}{c}{94} & \multicolumn{2}{c}{95} & \multicolumn{2}{c}{96} & \multicolumn{2}{c}{97} \\
        \midrule
        \ioqp & $17$,$423$ & (93.2$\times$) & $17$,$423$ & (84.6$\times$) & $18$,$808$ & (91.2$\times$) & $21$,$910$ & (98.7$\times$) & $24$,$382$ & (90.6$\times$) & $31$,$843$ & (105.1$\times$)& $35$,$735$ & (102.7$\times$) & $51$,$522$ & (97.0$\times$)\\
        \bruchetal
        &$4{,}169$ & (22.3$\times$) & $4{,}984$ & (24.2$\times$) & $6{,}442$ & (31.3$\times$) & $7{,}176$ & (32.3$\times$) & $8{,}516$ & (31.7$\times$) & $10{,}254$ & (33.8$\times$) & $12{,}881$ & (37.0$\times$) & $15{,}840$ & (29.8$\times$)
        \\
        \grassrma
        & $807$ & (4.3$\times$) & $867$ & (4.2$\times$) & $956$ & (4.6$\times$) & $1{,}060$ & (4.8$\times$) & $1{,}168$ & (4.3$\times$) & $1{,}271$ & (4.2$\times$) & $1{,}577$ & (4.5$\times$) & $1{,}984$ & (3.7$\times$) \\

        \pyann 
        & $489$ & (2.6$\times$) & $539$ & (2.6$\times$) & $603$ & (2.9$\times$) & $654$ & (2.9$\times$) & $845$ & (3.1$\times$) & $1{,}016$ & (3.4$\times$) & $1{,}257$ & (3.6$\times$) & $1{,}878$ & (3.5$\times$) \\

        \textbf{\our (ours)} & 187 & \multicolumn{1}{c}{--} & 206 & \multicolumn{1}{c}{--} & 206 & \multicolumn{1}{c}{--} & 222 & \multicolumn{1}{c}{--} & 269 & \multicolumn{1}{c}{--} & 303 & \multicolumn{1}{c}{--} & 348 & \multicolumn{1}{c}{--} & 531 & \multicolumn{1}{c}{--} \\
       
        \toprule
        \multicolumn{17}{c}{\esplade on \msmarco} \\
        \midrule
        \ioqp & 
        $7{,}857$ & (35.4$\times$) & $8{,}382$ & (37.8$\times$) & $8{,}892$ & (37.2$\times$) & $9{,}858$ & (41.2$\times$) & $10{,}591$ & (41.4$\times$) & $11{,}536$ & (30.7$\times$) & $11{,}934$ & (31.2$\times$) & $14{,}485$ & (24.9$\times$)
        \\
        \bruchetal & $4{,}643$ & (20.9$\times$) & $5{,}058$ & (22.8$\times$) & $5{,}869$ & (24.6$\times$) & $6{,}599$ & (27.6$\times$) & $7{,}555$ & (29.5$\times$) & $8{,}962$ & (23.8$\times$) & $10{,}414$ & (27.2$\times$) & $13{,}883$ & (23.9$\times$)
        \\
        \grassrma & 
        $2{,}074$ & (9.3$\times$) & $2{,}658$ & (12.0$\times$) & $2{,}876$ & (12.0$\times$) & $3{,}490$ & (14.6$\times$) & $4{,}431$ & (17.3$\times$) & $5{,}141$ & (13.7$\times$) & $7{,}181$ & (18.7$\times$) & $12{,}047$ & (20.7$\times$)
        \\
        \pyann &
        $1{,}685$ & (7.6$\times$) & $1{,}702$ & (7.7$\times$) & $2{,}045$ & (8.6$\times$) & $2{,}409$ & (10.1$\times$) & $3{,}119$ & (12.2$\times$) & $4{,}522$ & (12.0$\times$) & $7{,}317$ & (19.1$\times$) & $12{,}578$ & (21.6$\times$)
        \\
        \textbf{\our (ours)} & $222$ & \multicolumn{1}{c}{--} & $222$ & \multicolumn{1}{c}{--} & $239$ & \multicolumn{1}{c}{--} & $239$ & \multicolumn{1}{c}{--} & $256$ & \multicolumn{1}{c}{--} & $376$ & \multicolumn{1}{c}{--} & $383$ & \multicolumn{1}{c}{--} & $581$ & \multicolumn{1}{c}{--} \\

        \toprule
        \multicolumn{17}{c}{\unicoil on \msmarco} \\
        \midrule
        \ioqp & 
        $22{,}278$ & (193.7$\times$) & $25{,}060$ & (203.7$\times$) & $26{,}541$ & (199.6$\times$) & $30{,}410$ & (181.0$\times$) & $33{,}327$ & (198.4$\times$) & $34{,}061$ & (189.2$\times$) & $38{,}399$ & (143.3$\times$) & $40{,}759$ & (145.6$\times$)
        \\
        \bruchetal & $6{,}375$ & (55.4$\times$) & $7{,}072$ & (57.5$\times$) & $8{,}192$ & (61.6$\times$) & $9{,}207$ & (54.8$\times$) & $10{,}306$ & (61.3$\times$) & $12{,}308$ & (68.4$\times$) & $14{,}359$ & (53.6$\times$) & $17{,}572$ & (62.8$\times$)
        \\
        \grassrma & 
        $1{,}318$ & (11.5$\times$) & $1{,}434$ & (11.7$\times$) & $1{,}812$ & (13.6$\times$) & $2{,}004$ & (11.9$\times$) & $2{,}168$ & (12.9$\times$) & $2{,}668$ & (14.8$\times$) & $4{,}140$ & (15.4$\times$) & $5{,}340$ & (19.1$\times$)
        \\
        \pyann 
        & $1{,}133$ & (9.9$\times$) & $1{,}456$ & (11.8$\times$) & $1{,}741$ & (13.1$\times$) & $1{,}755$ & (10.4$\times$) & $2{,}061$ & (12.3$\times$) & $2{,}973$ & (16.5$\times$) & $3{,}883$ & (14.5$\times$) & $6{,}324$ & (22.6$\times$)
        \\
        \textbf{\our (ours)} & 115 & \multicolumn{1}{c}{--} & 123 & \multicolumn{1}{c}{--} & 133 & \multicolumn{1}{c}{--} & 168 & \multicolumn{1}{c}{--} & 168 & \multicolumn{1}{c}{--} & 180 & \multicolumn{1}{c}{--} & 268 & \multicolumn{1}{c}{--} & 280 & \multicolumn{1}{c}{--} \\
        
        \toprule
        \multicolumn{17}{c}{\splade on \nq} \\
        \midrule
        \ioqp 
        & $8{,}313$ & (42.6$\times$) & $8{,}854$ & (45.4$\times$) & $9{,}334$ & (44.2$\times$) & $11{,}049$ & (46.0$\times$) & $11{,}996$ & (48.0$\times$) & $14{,}180$ & (53.3$\times$) & $15{,}254$ & (53.3$\times$) & $18{,}120$ & (50.1$\times$)
        \\
        \bruchetal 
        & $3{,}862$ & (19.8$\times$) & $4{,}309$ & (22.1$\times$) & $4{,}679$ & (22.2$\times$) & $5{,}464$ & (22.8$\times$) & $6{,}113$ & (24.5$\times$) & $6{,}675$ & (25.1$\times$) & $7{,}796$ & (27.3$\times$) & $9{,}109$ & (25.2$\times$)
        \\
        \grassrma 
        & $1{,}000$ & (5.1$\times$) & $1{,}138$ & (5.8$\times$) & $1{,}208$ & (5.7$\times$) & $1{,}413$ & (5.9$\times$) & $1{,}549$ & (6.2$\times$) & $2{,}091$ & (7.9$\times$) & $2{,}448$ & (8.6$\times$) & $3{,}038$ & (8.4$\times$)
        \\
        \pyann 
        & $610$ & (3.1$\times$) & $668$ & (3.4$\times$) & $748$ & (3.5$\times$) & $870$ & (3.6$\times$) & $1{,}224$ & (4.9$\times$) & $1{,}245$ & (4.7$\times$) & $1{,}469$ & (5.1$\times$) & $1{,}942$ & (5.4$\times$)
        \\
        \textbf{\our (ours)} & 195 & \multicolumn{1}{c}{--} & 195 & \multicolumn{1}{c}{--} & 211 & \multicolumn{1}{c}{--} & 240 & \multicolumn{1}{c}{--} & 250  & \multicolumn{1}{c}{--} & 266  & \multicolumn{1}{c}{--} & 286  & \multicolumn{1}{c}{--} & 362 & \multicolumn{1}{c}{--} \\
        \bottomrule
    \end{tabular}}
	\caption{Mean latency ($\mu$sec/query) at different accuracy cutoffs with speedup (in parenthesis) gained by \our over others.\label{table:results1}}
    \vspace{-5mm}
\end{table*}

\vspace{1mm}
\noindent \textbf{Baselines}.
We compare \our with five state-of-the-art retrieval solutions. Two of these are the
winning solutions of the ``Sparse Track'' at the 2023 BigANN Challenge\footnote{\url{https://big-ann-benchmarks.com/neurips23.html}} at NeurIPS.
These include:
\begin{itemize}[leftmargin=*]
    \item \grassrma: A graph-based method for dense vectors adapted to sparse vectors that appears in the BigANN challenge as ``\textsc{sHnsw}.''\footnote{C++ code is publicly available at \url{https://github.com/Leslie-Chung/GrassRMA}.}
    \item \pyann: Another graph-based ANN solution.\footnote{C++ code is publicly available at \url{https://github.com/veaaaab/pyanns}.}
\end{itemize}
The other three baselines are inverted index-based solutions:
\begin{itemize}[leftmargin=*]
    \item \ioqp~\cite{mpg22-desires}: Impact-sorted query processor written in Rust. We choose \ioqp because it is known to outperform JASS~\cite{jass2015}, a widely-adopted open-source impact-sorted query processor.
    \item \bruchetal~\cite{bruch2023bridging}: An inverted index where lists are partitioned into blocks through clustering. At query time, after finding the $N$ closest clusters to the query, a coordinate-at-a-time
    algorithm traverses the inverted lists. The solution is approximate because only documents that belong to 
    top $N$ clusters are considered.
    \item \pisa~\cite{MSMS2019}: An inverted index-based C++ library based on \texttt{ds2i}~\cite{pefi-SIGIR14} that uses highly-optimized blocked variants of WAND.
    \pisa is \emph{exact} as it traverses inverted lists in a rank-safe manner.
\end{itemize}

We also considered the method by Lassance \etal~\cite{lassance2023static-pruning}.
Their approach statically prunes either inverted lists (by keeping $p$-quantile of elements),
documents (by keeping a fixed number of top entries), or all coordinates whose value is 
below a threshold. While simple,~\cite{lassance2023static-pruning} is only able to speed up query processing by 2--4$\times$
over \pisa on \esplade embeddings of \msmarco. We found it to be ineffective on
\splade and generally far slower than \grassrma and \pyann. As such we do not include it in our discussions.

We build \ioqp and \pisa indexes using Anserini\footnote{\url{https://github.com/castorini/anserini}} and
apply recursive graph bisection~\cite{mpm21-sigir}. For \ioqp, we vary the \emph{fraction} (of the total collection)
hyper-parameter in $[0.1, 1]$ with step size of $0.05$. For \bruchetal, we sketch documents using \sinnamon
and a sketch size of $1{,}024$, and build $4 \sqrt{N}$ clusters, where $N$ is the number of documents in the collection.
For \grassrma and \pyann, we build different indexes by running all possible combinations of
$ef_c \in  \{1000, 2000\}$ and $M \in \{16, 32, 64, 128, 256 \}$.
During search we test $ef_s \in [5, 100]$ with step size $5$, then $[100, 400]$ with step $10$, $[100, 1000]$ with step $100$, and finally $[1000, 5000]$ with step $500$.
We apply early stopping when accuracy saturates.

Our grid search for \our on \msmarco is over:
$\lambda \in [1500, 7500]$ with step size of $500$,
$\beta \in [150, 750]$ with step $50$,
and $\alpha \in [0.1, 0.5]$ with $0.1$. 
Best results are achieved with $\lambda=6{,}000$, $\beta=400$, and $\alpha=0.4$.
The grid search for \our on \nq is over: $\lambda \in \{4500, 5250, 6000\}$,
$\beta \in \{300, 350, 400, 450, 525, 600, 700, 800\}$,
and $\alpha \in \{0.3, 0.4, 0.5\}$. Best results are achieved with $\lambda=5{,}250$, $\beta=525$, and $\alpha=0.5$. \our employs 8-bit scalar quantization for summaries. Moreover, \our uses matrix multiplication to efficiently multiply the query vector with all quantized summaries of an inverted list.

\vspace{1mm}
\noindent \textbf{Evaluation Metrics}.
We evaluate all methods using three metrics:
\begin{itemize}[leftmargin=*]
    \item Latency ($\mu$sec.). The time elapsed, in \emph{microseconds}, from the moment a query vector is presented
    to the index to the moment it returns the requested top $k$ document vectors running in single thread mode.
    Latency does not include embedding time.
    \item Accuracy. The percentage of true nearest neighbors recalled in the returned set.
    By measuring the recall of an approximate set given the exact top-$k$ set,
    we study the impact of the different levers in an algorithm on its overall accuracy as
    a retrieval engine.
    \item Index size (MiB). The space the index occupies in memory.
\end{itemize}

\vspace{1mm}
\noindent \textbf{Reproducibility and Hardware Details}.
We implemented \our{} in Rust.\footnote{Our code is publicly available at \url{https://github.com/TusKANNy/seismic}.}
We compile \our{} by using the version $1{.}77$ of Rust and use the highest level of
optimization made available by the compiler.
We conduct experiments on a server equipped with one Intel i9-9900K CPU with
a clock rate of $3{.}60$ GHz and $64$ GiB of RAM.
The CPU has $8$ physical cores and $8$ hyper-threaded ones. We query the index using a single thread.

\subsection{Results}
We now present our experimental results. We begin by comparing the performance of \our with baselines.
We then ablate \our to understand the impact of our design choices on performance.

\subsubsection{Accuracy-Latency Trade-off}
Table~\ref{table:results1} details retrieval performance in terms of average per-query latency
for \splade, \esplade, and \unicoil on \msmarco, and \splade on \nq.
We frame the results as the trade-off between effectiveness and efficiency. In other words,
we report mean per-query latency at a given accuracy level.

The results on these datasets show \our's remarkable relative efficiency,
reaching a latency that is often one to two orders of magnitude smaller.
Overall, \our consistently outperforms all baselines at all accuracy levels, including
\grassrma and \pyann, which in turn perform better than other inverted index-based
baselines---confirming the findings of the BigANN Challenge.

We make a few additional observations.
\ioqp appears to be the slowest method across datasets. This is not surprising considering
the distributional abnormalities of learned sparse vectors, as discussed earlier.
\bruchetal generally improves over \ioqp, but \our speeds up query processing further.
In fact, the minimum speedup over \ioqp (\bruchetal) on \msmarco is
$84.6\times$ ($22.3\times$) on \splade, $24.9\times$ ($20.9\times$) on \esplade,
and $143.3\times$ ($53.6\times$) on \unicoil.

\our consistently outperforms \grassrma and \pyann
by a substantial margin, ranging from $2.6\times$ (\splade on \msmarco) to $21.6\times$
(\esplade on \msmarco) depending on the level of accuracy.
In fact, as accuracy increases, the latency gap between \our
and the two graph-based methods widens.
This gap is much larger when query vectors are sparser, such as with \esplade embeddings.
That is because, when queries are highly sparse, inner products between queries and documents
become smaller, reducing the efficacy of a greedy graph traversal.
As one data point, \pyann over \esplade embeddings of \msmarco visits roughly
$40{,}000$ documents to reach $97\%$ accuracy, whereas \our evaluates just $2{,}198$ documents.

Finally, we highlight that \pisa is the slowest (albeit, \emph{exact}) solution.
On \msmarco, \pisa processes queries in about $100{,}325$ microseconds on \splade
embeddings. On \esplade and \unicoil, its average latency is
$7{,}947$ and $9{,}214$ microseconds, respectively. We note that its
high latency on \splade is largely due to the much larger number of
non-zero entries in queries.

\begin{figure}[t]
	\centering
	\includegraphics[width=0.9\columnwidth]{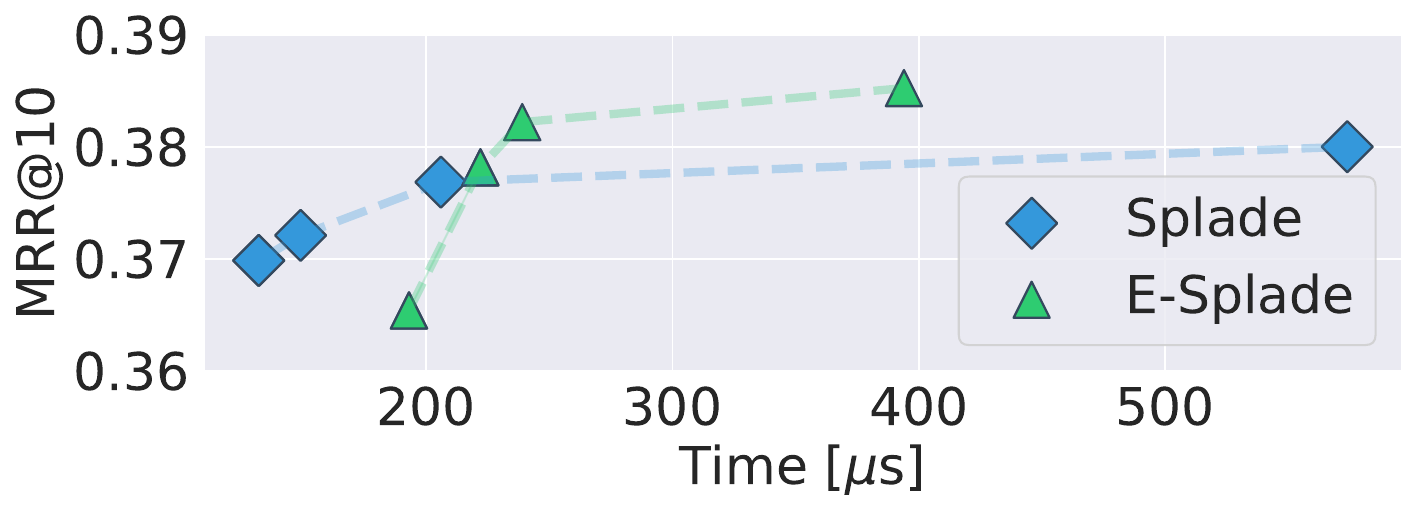}
	\caption{MRR@10 on \msmarco.\label{fig:msmarco-mrr-plot}}
\end{figure}

We conclude with a remark on the relationship between retrieval accuracy
(as measured by recall with respect to exact search) and ranking quality (such as MRR and
NDCG~\cite{Jarvelin2002} given relevance judgments). Even though ranking quality is not
our primary focus, we measured MRR@10 on \msmarco for the approximate top-$k$ sets obtained from \our,
and plot that as a function of per-query latency in Figure~\ref{fig:msmarco-mrr-plot}.
While MRR@10 is relatively stable, we do notice a drop in the low-latency (and thus low-accuracy) regime.
Perhaps more interesting is the fact that \our can speed up retrieval over \splade so much that
if the time budget is tight, using \splade embeddings gets us to a higher MRR@10 faster.

\subsubsection{Space and Build Time}
Table~\ref{table:results-rq2} records the time it takes to index the entire \msmarco collection embedded with \splade with different methods, and the size of the resulting index. We perform this experiment on a machine with two Intel Xeon Silver 4314 CPUs clocked at 2.40GHz, with 32 physical cores plus 32 hyper-threaded ones and 512 GiB of RAM. We build the indexes by using multi-threading parallelism with $64$ cores.

We left out the build time for \ioqp because its index construction has many external dependencies
(such as Anserini and graph bisection) that makes giving an accurate estimate difficult.

Trends for other datasets are similar to those reported in Table~\ref{table:results-rq2}.
Notably, indexes produced by approximate methods are larger.
That makes sense: using more auxiliary statistics helps narrow the search space dynamically
and quickly. Among the highly efficient methods, the size of \our's index is mild,
especially compared with \grassrma. Importantly, \our builds its index in a fraction of
the time it takes \pyann or \grassrma to index the collection.

\begin{table}[t]
	\centering
\adjustbox{max width=\textwidth}{%
	\begin{tabular}{lrr}
    \toprule
    \multicolumn{3}{c}{\splade on \msmarco} \\
    \midrule
    Model & Index size (MiB) & Index build time (min.) \\
    \midrule
    \ioqp & $2{,}195$ & - \\
    \bruchetal & $8$,$830$ & $44$\\
    \grassrma & $10$,$489$ &$267$ \\
    \pyann & $5$,$262$ & $137$\\
    \textbf{\our (ours)} & $6$,$416$ &  $5$\\
    \bottomrule
	\end{tabular}}
	\caption{Index size and build time.\label{table:results-rq2}}
    \vspace{-5mm}
\end{table}

\subsection{Ablation Study}
We now take \our apart to study the impact of its components.
We take the \splade embeddings of \msmarco and analyze the impact of (a) quantization on summaries;
(b) two strategies to partition inverted lists; and
(c) two methods for building the summary vectors.

\vspace{1mm}
\noindent 
\textbf{Quantization of Summaries}.
We empirically observe that the scalar quantization applied to summaries does not hinder the effectiveness or the efficiency of \our. Indeed, it reduces the memory footprint of the summaries by a factor of $4$.

\begin{figure}[t]
    \centering
    \includegraphics[width=0.9\columnwidth]{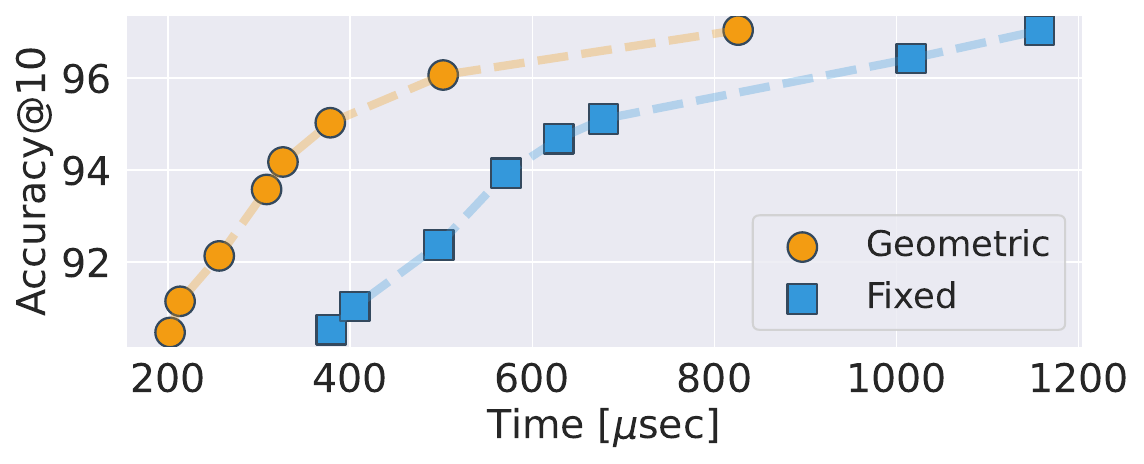}
    \caption{Fixed vs. geometric blocking. Data sampled from parameters: $\cut \in \{1,\ldots, 10\}$
    and $\heapfactor \in \{0.7, 0.8, 0.9, 1.0\}$.\label{fig:variable_vs_fixed}}
\end{figure}

\vspace{1mm}
\noindent \textbf{Fixed vs. Geometric Blocking}.
We delegate inverted list blocking to a clustering algorithm.
In this section, we wish to understand the impact of \emph{geometric} clustering on the performance of \our.
To that end, we compare two realizations of the index.
In one, called ``geometric'' blocking, we use a variant of K-Means as described in Section~\ref{sec:methodology:blocking}.
Separately, in what we call ``fixed'' blocking, we take the impact-sorted inverted lists and chunk them into
fixed-size groups. We then compare the performance of these two configurations on the accuracy-latency trade-off space.
Figure~\ref{fig:variable_vs_fixed} reports our results, showing that geometric blocking significantly
outperforms fixed blocking for all ranges of hyper-parameters considered.

\vspace{1mm}
\noindent \textbf{Fixed vs. Importance-based Summaries}.
Recall that, our summary vectors are $\alpha$-mass subvectors of the vector produced by
Equation~(\ref{equation:summary}). In a sense, the summary reflects the distribution of documents within a block.
Here, we contrast that ``importance-based'' summary generation with a simple alternative:
Keeping a \emph{fixed} number of top entries of the vector from Equation~(\ref{equation:summary}).
The drawback of this alternative is that we store the same number of entries
for each block regardless of the number of documents in the block or the distribution of their importance, thus weakening the performance of \our.

\begin{figure}[t]
    \centering
    \includegraphics[width=0.9\columnwidth]{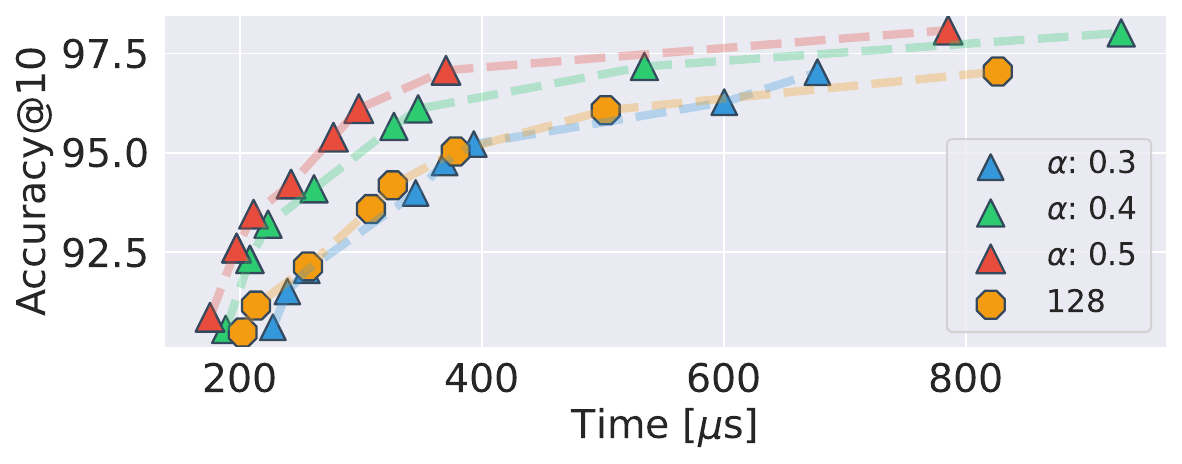}
    \caption{Fixed ($128$ top entries per summary) vs. importance-based ($\alpha$-mass subvectors) summaries.\label{fig:summary_energy}}
\end{figure}

Figure \ref{fig:summary_energy} visualizes the latency-accuracy trade-off of these different settings.
It is clear that, for a fixed time budget, importance-based summaries lead to better accuracy than
fixed-length summaries. Moreover,
summaries with $128$ top entries take $2{,}687$ MiB of space,
while importance-based summaries with $\alpha=0.5$ consume $2{,}885$ MiB (without quantization).
Reducing $\alpha$ to $0.4$ and $0.3$ lowers the size to $2{,}303$ and $1{,}801$ MiB, respectively. 

\vspace{1mm}
\noindent \textbf{Forward Index}.
The forward index could use $32$- or $16$-bit floating points to store vector values.
We use half-precision, leading to $4{,}113$ MiB of space usage at negligible cost
to accuracy and no impact on latency. We confirm that \pyann too uses this representation.

\begin{figure}[t]
	\centering
	\includegraphics[width=0.9\columnwidth]{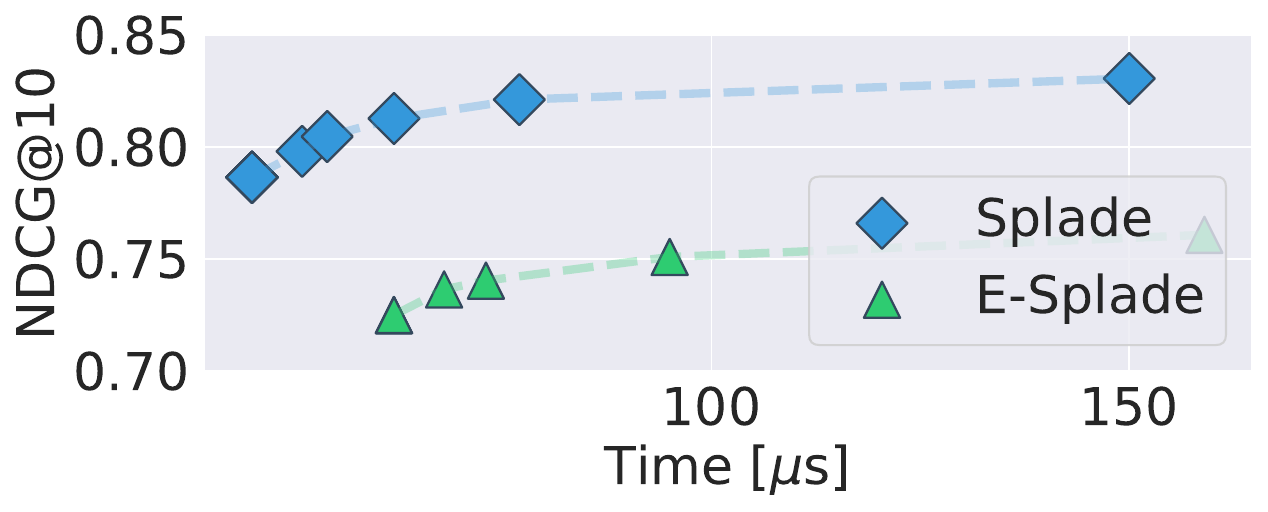}
	\caption{NDCG@10 on the \quora dataset.\label{fig:beirplot}}
\end{figure}

\section{Concluding Remarks}
\label{sec:conclusions}
We presented \our, a novel approximate algorithm that facilitates effective and efficient retrieval over learned sparse embeddings.
We showed empirically its remarkable efficiency on a number of embeddings of publicly-available datasets.
\our outperforms existing methods, including the winning, graph-based algorithms at the BigANN Challenge in NeurIPS 2023
that use similar-sized (or larger) indexes.

One of the exciting opportunities that our research creates is that it offers a new way of thinking
about sparse embedding models. Let us explain how. When \splade proved difficult to scale because
state-of-the-art inverted index-based solutions failed to process queries fast enough, the community
moved towards \esplade and other variants that reduce query processing time, but that exhibit degraded performance
in zero-shot settings. Evidence suggests, for example, that \esplade embeddings of \quora---a \beir dataset---yield
NDCG@10 of $0.76$ while \splade embeddings yield $0.83$.

\our changes that equation. As we visualize in Figure~\ref{fig:beirplot}, for any given time budget,
\our retrieves a better-quality top-$k$ set from the \splade embeddings of \quora.
The key take-away message is clear: \our speeds up retrieval over \splade so dramatically that
switching to \esplade becomes unnecessary and, in fact, detrimental to both efficiency and effectiveness.

As future work,
we intend to explore the application of compression techniques
for inverted lists~\cite{CSUR21} to further reduce the size of inverted and forward indexes.

\vspace{1mm}
\noindent \textbf{Acknowledgements}.
This work was partially supported by the Horizon Europe RIA ``Extreme Food Risk Analytics'' (EFRA), grant agreement n. 101093026, by the PNRR - M4C2 - Investimento 1.3, Partenariato Esteso PE00000013 - ``FAIR - Future Artificial Intelligence Research'' - Spoke 1 ``Human-centered AI'' funded by the European Commission under the NextGeneration EU program, by the PNRR ECS00000017 Tuscany Health Ecosystem Spoke 6 ``Precision medicine \& personalized healthcare'' funded by the European Commission under the NextGeneration EU programme, by the MUR-PRIN 2017 ``Algorithms, Data Structures and Combinatorics for Machine Learning'', grant agreement n. 2017K7XPAN\_003, and by the MUR-PRIN 2022 ``Algorithmic Problems and Machine Learning'', grant agreement n. 20229BCXNW.

\bibliographystyle{ACM-Reference-Format}
\bibliography{biblio}

\end{document}